\documentclass[12pt,a4paper]{article}
\begin{document}

\title{Canonical LQG operators and kinematical states for plane gravitational waves}
\author{F. Hinterleitner,\\Department of theoretical physics and astrophysics,\\
 Masaryk university Brno, Czech Republic}
\maketitle \abstract{In a 1+1 dimensional model of plane
gravitational waves the flux-holonomy algebra of loop quantum
gravity is modified in such a way that the new basic operators
satisfy canonical commutation relations. Thanks to this construction
it is possible to find kinematical solutions for unidirectional
plane gravitational waves with finite geometric expectation values
and fluctuations, which was problematic in a more conventional
approach in a foregoing paper by the author and coauthors
\cite{adel}.}
\section{Introduction}
Nonperturbative canonical quantum gravity comes in two steps: The
first one is a formulation of general relativity in terms of
connection and triads on a spacelike hypersurface - the Ashtekar
variables - where the total Hamiltonian is a combination of
constraints. The constraints form a first-class Poisson bracket
algebra.

In the second step quantum operators and states are constructed. In
this process, not the connection components themselves, but their
holonomies play the role of fundamental variables. Thus, before
promoting the constraints or other functions of the connection to
operators, the connection has to be reformulated in terms of
holonomies, in such a way that for weak gravitational fields and in
the continuous limit the original formulations are approximated.
This leads to the problem that the Poisson bracket algebra of
constraints does not carry over identically to the commutator
algebra of the corresponding constraint ope\-rators. The present
approach to a simplified 1+1 dimensional model is guided by two
principles:

1) We construct slightly modified operators following the prototypes
in ele\-mentary quantum mechanics, with configuration variables
promoted to multiplication operators and conjugate momenta to
derivatives. In loop quantum gravity (LQG) state functions are
functions of group elements (holonomies), so we introduce in section
3 as fundamental operators multiplication by group elements in the
fundamental representation and derivative operators with respect to
them, instead of derivatives with respect to Lie algebra elements.
In this way the fundamental operators commute canonically.

2) In LQG eigenvalues of triad operators usually have both signs,
which leads, in contrast to classical theory, to identical copies of
the metric geometry with different orientations of spatial
directions. It is natural that quantum operators, as far as they are
not related to spatial orientation, should act in an equivalent way
in sectors of geometry differing only by orientation. This leads to
slightly different, but quite natural constructions of corresponding
operators in different sectors. In the calculations in section 4 it
turns out that such a choice is necessary for physically acceptable
results in all sectors.

The model, which our attention is directed to in this paper, is a
model of plane gravitational waves \cite{plane}, derived from a
Gowdy model formulated in Ashtekar variables in \cite{BD1,BD2}.
Being homogeneous in two directions, this is an example of an
effectively 1+1 dimensional midi-superspace. In this model the new
construction of canonically commuting operators is applied to the
formulation of a unidirectionality constraint operator and its
solutions. Another interesting approach to models of this type with
a modification of operators is the abelianization of the Hamiltonian
constraint \cite{blas},

\section{The model}
In the model of plane gravitational waves, the physical object of
this paper, we assume homogeneity in the $(x,y)$ plane and
propagation in the $z$ direction. As a further simplification we
assume linearly polarized waves. The Ashtekar variables are the
following: Connection components $X$, $Y$ in the $x$ and $y$
direction and $\cal A$ in the $z$ direction, and respective
conjugate densitized triads $E^x$, $E^y$, and $\cal E$. On a
spacelike hypersurface all these variables depend only on $z$. In
terms of these variables the spatial metric has the form
\begin{equation}\label{me}
{\rm d}s^2={\cal E}\frac{E^y}{E^x}\,{\rm d}x^2+{\cal
E}\frac{E^x}{E^y}\,{\rm d}y^2+\frac{E^xE^y}{\cal E}\,{\rm d}z^2.
\end{equation}
The Gau\ss, diffeo, and Hamiltonian constraint of the system are
given in \cite{BD1,BD2}.

The graph $G$, on which one-dimensional analogs of spin networks
(SNW) are defined, is the $z$ axis, divided into a sequence of links
$l_i$ by nodes $n_i$ at the locations $z_i$. In \cite{BD2} basic
quantum state functions are constructed from the point holonomies
$\exp(i\frac{\mu_i}{2}\,X(z_i))$ and
$\exp(i\frac{\nu_i}{2}\,Y(z_i))$ at the nodes and the link
holonomies $\exp\left(i\frac{k_i}{2}\int_{\ell_i}{\cal A}\right)$.
The point holonomies lie in $\bf R_{\rm Bohr}$, the group of the
Bohr compactification of the reals, link holonomies are $U(1)$
functions. The combined state functions (for convenience without the
factors of 1/2 in the exponents present in \cite{BD2}) are
\begin{equation}\label{state}
\prod_{\ell_j\in G}\exp\left(i\,k_j\int_{\ell_j}{\cal
A}\right)\prod_{n_i\in
N(G)}\exp\left(i\,\mu_i\,X(n_i)\right)\;\exp\left(i\,\nu_i\,Y(n_i)\right),
\end{equation}
where $N(G)$ denotes the set of nodes. In the following we
concentrate on one node, and on one or two links, and denote point
holonomies by $|\mu,\nu\rangle$ and link holonomies by $|k\rangle$,
omitting the indices $i$ and $j$.

In \cite{BD2} holonomy operators acting on point holonomies are
defined as $SU(2)$ operators, in their action on the states traces
of $SU(2)$ generators have to be taken. In the following we take the
$U(1)$ operators
\begin{equation}
\hat U_x=\exp(i\,X),
\end{equation}
and $\hat U_y$ analogously, for simplicity and for a more natural
action on the functions (\ref{state}). In their action on arbitrary
nodes they raise the labels $\mu$ and $\nu$ in (\ref{state}) by one.
As indicated above, in \cite{BD2} the point holonomies are
introduced as unitary representations of ${\bf R}_{\rm Bohr}$  and
holonomies as operators shifting the labels of these
representations. Later on, it will turn out that only states
$|m,n\rangle$ with integer labels $\mu=m$ and $\nu=n$ are of
interest. In the solutions of our model only such series of states
out of the representations of ${\bf R}_{\rm Bohr}$ contain states
with $m$ or $n$ or both being equal to zero. With such a reduction
also point holonomies lie in $U(1)$. In section 4 this series will
be distinguished as possible sets of kinematical states.

As in \cite{BD2}, the states $|k\rangle$ are considered to lie in
$U(1)$. The holonomy operator
\begin{equation}\label{5}
\hat{\cal U}_\ell=\exp\left(i\int_\ell{\cal A}\right)
\end{equation}
multiplies state functions by an element of the fundamental
representation of $U(1)$ and as such it raises the label $k$ of the
representation of the state function $|k\rangle$ by one.

The densitized triads $E^x$ and $E^y$ are scalar densities, when
integrated over some interval $I$ on the $z$ axis, they give rise to
flux operators with nontrivial action when there is a node in $I$.
Then the action of the operators $\bar{E^x}=-i\delta/\delta X$ and
$\bar{E^y}=-i\delta/\delta Y$ on a node function is
\begin{equation}
\int_I\bar {E^x}\:|m,n\rangle=m\,|m,n\rangle, \hspace{2cm}
\int_I\bar{E^y}\:|m,n\rangle=n\,|m,n\rangle
\end{equation}
up to a factor containing the square of the Planck length which we
set equal to one. These operators, taken over from \cite{BD2}, are
denoted by a bar instead of the usual hat, as we will introduce
different operators for the same quantities in the next section.

${\cal E}(z)$, as a scalar, acts directly at the point $z$. Up to
the mentioned type of factor the action is
\begin{equation}
\bar{\cal E}(z)\,|k\rangle=k\,|k\rangle,
\end{equation}
when $z$ lies on a link with label $k$. The meaning of ${\cal E}(z)$
is the geometrical area of a plane of unit coordinate area,
transversal to a link, i.\,e. a cross-section area of the
gravitational wave. When at $z$ there is a node and when $k_-$ and
$k_+$ are the labels of the link functions left and right from $z$
and $|\psi\rangle$ is a SNW function containing $|k_+\rangle$ and
$|k_-\rangle$ then
\begin{equation}
\bar{\cal E}(z)\;|\psi\rangle=\frac{k_++k_-}{2}\:|\psi\rangle.
\end{equation}
All these triad operators act diagonally on state functions in the
SNW basis.

\section{Redefinition of basic operators}
In the foregoing section we have briefly introduced the triad
operators as in \cite{BD1} and holonomy operators in a simplified
form (from $SU(2)$ to $U(1)$) which, nevertheless, is sufficient for
acting on $U(1)$ state functions.

A slightly modified construction of operators starts from the fact
that the variables to describe quantum states are functions of group
elements, namely $U(1)$ holonomies. Let's take a point holonomy of
$X$ (the construction of the $Y$ holonomies is analogous)
\begin{equation}\label{1}
U_x(z)=e^{i m X(z)}.
\end{equation}
For every node, group elements are labeled by a number $X(z)$ on the
manifold of $U(1)$ - a circle - with $0\leq X(z) < 2\pi$. Integers
$m$ label irreducible representations, for $m=1$ we have the
fundamental one, lets denote it by
\begin{equation}\label{2}
g(z)=e^{iX(z)}.
\end{equation}
$X(z)$ is a local generator on the one-dimensional space manifold.
Quantum state functions at nodes are functions on $U(1)\times U(1)$,
a basis for the $X$ functions is given by the point holonomies
\begin{equation}
U_x(z)=e^{im X(z)}=g^m.
\end{equation}

The basic idea for a modified construction of operators on this
space of functions is to replace $X$ and $E^x$ by canonical
variables in terms of holonomies:
\begin{equation}\label{11}
X\rightarrow U_x-{\bf1}, \hspace{5mm} E^x\rightarrow
-i\frac{\delta}{\delta U_x}=-i\frac{\delta X}{\delta
U_x}\,\frac{\delta}{\delta X}=-U_x^{-1}\frac{\delta}{\delta X},
\end{equation}
$\bf1$ is the unit operator. The multiplication operator $\hat
U_x(z)$ multiplies the state function at $z$ by the holonomy
(\ref{2}), in other words, it raises the label $m$ by one,
\begin{equation}
\hat U_x(z)=e^{i X(z)}, \hspace{1cm} \hat
U_x(z)\,|m,n\rangle=|m+1,n\rangle.
\end{equation}

The derivative operator $i\delta/\delta(X)$ does not commute
canonically with $\hat U_x$, the commutator is a holonomy. A
canonically conjugate operator to $\hat U_x$ is taken from
(\ref{11})
\begin{equation}\label{E1}
\hat E^x := -\hat U_x^{-1}(z)\,\delta/\delta X(z),\hspace{1cm}\hat
E^x(z)|m,\,n\rangle=-im\,|m-1,\,n\rangle,
\end{equation}
a lowering operator combined with a multiplier by $m$. The
commutator is
\begin{equation}
[\hat U_x(z_i),\hat E^x(z_j)]=i\,\delta_{ij}.
\end{equation}
when $z_i$ and $z_j$ are the coordinates of nodes.

When we formulate an operator corresponding to $X$ in the form
\begin{equation}\label{U1}
\hat X(z) := \hat U(z) - \bf1,
\end{equation}
it is, of course, also canonically conjugate to $\hat E$,
\begin{equation}\label{comm}
[\hat X(z_i),\hat E^x(z_j)]=i\,\delta_{ij}.
\end{equation}
$U_x-1$ is a good approximation in first order for small values of
$X(z)$, i.e. for weak fields, when we expect quantum theory to
approach the classical limit. The classical expression $-U_x^{-1}
E^x$ approximates the conventional differential operator
$\delta/\delta X$ in zeroth order in the limit of small $X$.

Alternatively we can define
\begin{equation}\label{tilde}
\tilde E^x := -\hat U_x(z)\,\delta/\delta X(z),\hspace{10mm}\tilde
E^x|m,n\rangle=-im\,|m+1,n\rangle
\end{equation}
and replace $\hat X$ by
\begin{equation}\label{U2}
\tilde X(z):={\bf1}-\hat{U_x}^{-1}(z)
\end{equation}
with the same commutation relation as (\ref{comm}),
\begin{equation}
[\tilde X(z_i),\tilde E^x(z_j)]=i\,\delta_{ij}.
\end{equation}

From the local variables $\cal A$ and $\cal E$ we construct
operators acting on link holonomies. The holonomy operator (\ref{5})
raises the label $k$ of a link by 1,
\begin{equation}
\hat{\cal U}_\ell\,|k\rangle=|k+1\rangle.
\end{equation}
The triad operator $\hat{\cal E}_\ell$ is constructed from
\begin{equation}\label{triad}
{\cal E}_\ell=-{\cal U}_\ell^{-1}\,\frac{\delta}{\delta{\cal A}(z)},
\hspace{5mm}z\in\ell,
\end{equation}
or as alternative version $\tilde{\cal E}$ analogously to
(\ref{tilde}), with the respective actions on link functions
\begin{equation}\label{calE}
\hat{\cal E}_\ell\,|k\rangle=-ik\,|k-1\rangle,\hspace{15mm}
\tilde{\cal E}_\ell\,|k\rangle=-ik\,|k+1\rangle.
\end{equation}
The operators constructed from the connection component $\cal A$ are
\begin{equation}\label{25}\hat{\cal A}:=\hat{\cal U}-\bf1 \hspace{5mm}
\mbox{or} \hspace{5mm} \tilde {\cal A}:={\bf1}-\hat{\cal U}^{-1},
\end{equation}
their corresponding classical expressions are good approximations
for $\cal A$ in first order for short links. The commutators with
the triad operators are canonical
\begin{equation}\label{commutator}
[\hat{\cal A}_\ell,\,\hat{\cal E}_{\ell'}]=[\tilde{\cal
A}_\ell,\,\tilde{\cal E}_{\ell'}]=i\,\delta(\ell,\ell').
\end{equation}
$\delta(\ell,\ell')$ is one if $\ell$ and $\ell'$ are the same link,
otherwise zero. In the continuous limit (\ref{triad}) approaches a
mere derivative operator, as for short links $\cal U$ and ${\cal
U}^{-1}$ approach identity in zeroth order.

\section{\bf The Killing constraint for unidirectional waves}

In \cite{plane} the condition of unidirectionality of plane
gravitational waves was formulated in form of first class
constraints to be imposed in addition to the constraints of
canonical general relativity. They are derived from the existence of
a null Killing field in the direction of wave propagation and have
the form
\begin{equation}\label{K}
K_\pm:=XE^x+YE^y\pm{\cal E}'.
\end{equation}
The prime denotes the derivative with respect to $z$, the expression
is of density weight one. The physical meaning is the following:
When the spatial metric (\ref{me}) is supplemented by a time
component $g_{tt}=-g_{zz}$ and zero shift vector, then the classical
expression
\begin{equation}
K_1:=XE^x+YE^y
\end{equation}
is the time derivative of $\cal E$ (see \cite{BD1}), thus
\begin{equation}
K_\pm=\dot{\cal E}\pm{\cal E}'.
\end{equation}
$K_+=0$ determines thus waves going into the positive $z$ direction
at the speed of light (called right-moving waves) and $K_-=0$
describes left-moving waves.

In \cite{adel} we attempted to quantize this constraint by
expressing the first part, $K_1$, in terms of a commutator of part
of the Hamiltonian constraint with the volume operator. Classically
$K_1=0$ and ${\cal E}'=0$ together distinguish a state without waves
at all, a one-dimensional description of the Minkowski vacuum. As
solutions of the corresponding quantum constraint equations we found
states that are normalizable, but, with the exception of the
zero-volume node state $|0,0\rangle$, they have diverging
expectation values of the length $\sqrt{E^xE^y/{\cal E}}$ between
two nodes and the volume $\sqrt{{\cal E}E^xE^y}$ associated to a
node. The situation becomes better when the classical constraint is
multiplied by some power of the volume and quantized afterwards. For
higher powers the convergence of length expectation values and
fluctuations become increasingly better, but this approach contains
an element of arbitrariness - which power should one choose?
Moreover, the constraints constructed in this way have different
density weights, as the volume is the determinant of the spatial
metric.

\subsection{Node operators and functions}
To obtain a real action of the operator $\hat K_+$, we multiply
(\ref{K}) by $i$ before defining an operator. Then with the choice
(\ref{E1}) and (\ref{U1}) for $(X,E^x)$ and analogously for
$(Y,E^y)$ the Killing constraint $K_+$ acquires the form
\begin{equation}\label{31}
\hat K_+(z):=i\hat X(z)\hat E^x(z) + i\hat Y\hat E^y(z) + i{\cal
E}'(z).
\end{equation}
In this subsection we anticipate eigenfunctions of $\hat{\cal E}$
(or $\tilde{\cal E}$) on the links left and right from $z$, so that
${\cal E}'(z)$ is the difference of eigenvalues, simply an imaginary
number (because of the $i$ in (\ref{calE}). A consistent application
of the operator $\hat{\cal E}/\tilde{\cal E}$ will be the given in
subsection (4.2). Preliminarily we define ${\cal D}=i{\cal E}'$ with
real $\cal D$. The action on a node function $|m,n\rangle$ is then
\begin{equation}
\hat K_+|m,n\rangle=({\cal
D}+m+n)|m,n\rangle-m\,|m-1,n\rangle-n\,|m,n-1\rangle.
\end{equation}
$\hat K_+$ contains lowering operators, acting from some state
$|m,n\rangle$ with positive $m$ and $n$ into the direction of the
$m$ and $n$ axes. When an axis is reached, the creation of new
states does not continue beyond it, due to the factors $m$ and $n$,
so the solutions have a finite number of nonvanishing coefficients.
Here it is essential that $|m,0\rangle$ and $|0,n\rangle$ are among
the solutions, otherwise the solutions would have an infinite number
of states $|m,n\rangle$ and diverging geometric expectation values.
This justifies the choice of integer labels in  physically relevant
node states. The resulting equation for the coefficients $a_{m,n}$
of the states $|m,n\rangle$ in an eigenstate
\begin{equation}\label{D}
|{\cal D}\rangle=\sum_{m,n}a_{m,n}\,|m,n\rangle
\end{equation}
of
\begin{equation}\label{hatk1}
\hat K_1:=\hat X\hat E^x+\hat Y\hat E^y
\end{equation}
with eigenvalue $-{\cal D}\geq0$ is the following:
\begin{equation}
({\cal D}+m+n)\,a_{m,n}-(m+1)\,a_{m+1,n}-(n+1)\,a_{m,n+1}=0.
\end{equation}
Consider first nonpositive integer values of ${\cal D}$:\\

\noindent{\bf Case 1}. Solutions in the first quadrant of the
$(m,n)$ plane, $m\geq0$, $n\geq0$:
\begin{enumerate}
\item ${\cal D}=0$: Here the only finite solution is $|0,0\rangle$.
\item ${\cal D}=-1$: There are two solutions,
\begin{equation}
\frac{1}{\sqrt{2}}\,(|0,0\rangle-|1,0\rangle) \hspace{5mm}
\mbox{and} \hspace{5mm} \frac{1}{\sqrt{2}}(|0,0\rangle-|0,1\rangle).
\end{equation}
\item ${\cal D}=-2$: Three finite solutions,
\begin{eqnarray}
\frac{1}{\sqrt{6}}\,(|0,0\rangle-2\,|1,0\rangle+|2,0\rangle),
\hspace{1cm}
\frac{1}{\sqrt{6}}\,(|0,0\rangle-2\,|0,1\rangle+|0,2\rangle)\nonumber\\
\mbox{and} \hspace{5mm}  \frac{1}{2}(|0,0\rangle - |1,0\rangle -
|0,1\rangle + |1,1\rangle).
\end{eqnarray}
Here appears the first nonzero expectation value
$\langle\sqrt{mn}\rangle=\frac{1}{4}$, and fluctuation
$\Delta(\sqrt{mn})=\frac{\sqrt{3}}{4}$ of the node contribution
$\sqrt{E^xE^y}$ to length and volume.

For larger negative values of ${\cal D}$ a pattern of binomial
coefficients appears. For unnormalized states, with $a_{0,0}=+1$ by
convention, we find:

\item ${\cal D}=-3$: A state with $n=0$ and the coefficients
$$a_{0,0}=1, \; a_{1,0}=-3,\; a_{2,0}=3, \; a_{3,0}=-1,$$
and one containing $n=0$ and $n=1$ and the coefficients
$$\begin{array}{lll}a_{0,0}=1,& a_{1,0}=-2,& a_{2,0}=1,\\
a_{0,1}=-1,& a_{1,1}=2,& a_{2,1}=-1\end{array}$$ and two further
states with $m$ and $n$ exchanged.
\item ${\cal D}=-4$: For $n=0$ the coefficients are
$$a_{0,0}=1, \; a_{1,0}=-4,\;  a_{2,0}=6, \; a_{3,0}=-4, \; a_{4,0}=1.$$
Then there is a state with $n=0$ and $n=1$ and
$$\begin{array}{llll}a_{0,0}=1,& a_{1,0}=-3,& a_{2,0}=3,& a_{3,0}=-1,\\
a_{0,1}=-1,& a_{1,1}=3,& a_{2,1}=-3,& a_{3,1}=1,\end{array}$$ and
finally a state with $n=0,1,$ or 2:
$$\begin{array}{lll}
a_{0,0}=1,& a_{1,0}=-2,& a_{2,0}=1,\\
a_{0,1}=-2,& a_{1,1}=4,& a_{2,1}=-2,\\
a_{0,2}=1, &a_{1,2}=-2,& a_{2,2}=1, \end{array}$$ and two further
states with $m\leftrightarrow n$.
\end{enumerate}

One can read off that for each negative integer ${\cal D}$ there are
$-{\cal D}+1$ solutions with $0\leq m\leq m_{\rm max}$ and $0\leq
n\leq n_{\rm max}$, such that $m_{\rm max}+n_{\rm max}=-{\cal D}$.
The general form of the unnormalized coefficients is
\begin{equation}
a_{m,n}=(-1)^{m+n}\left(\begin{array}{c}m_{\rm max}\\
m\end{array}\right)\left(\begin{array}{c}n_{\rm max}\\ n
\end{array}\right).
\end{equation}

\noindent{\bf Case 2}. $m\leq 0$, $n\leq0$: Here the states
$|m-1,n\rangle$ and $|m,n-1\rangle$, which would be created by the
above version of the operator from a state $|m,n\rangle$, lie
farther away from the axes than $|m,n\rangle$, so this operator
would create an infinity of states with a diverging expectation
value of $\sqrt{mn}$. It is the second version, according to
(\ref{tilde}) and (\ref{U2}) that acts in this case analogously to
the first version in case 1. This can be also expected for reasons
of symmetry: As the geometry of $|m,n\rangle$ and $|-m,-n\rangle$ is
the same up to the orientation of axes, the operator should act on
them in some analogous way, according to what was announced as
``principle 2" in the introduction.

Here the Killing operator is explicitly (the following equation
defines the operator $\tilde K_1$)
\begin{equation}\label{K1t}
\tilde K_+(z)=\tilde{K_1}(z)+i{\cal E}(z)':=i\tilde X(x)\tilde
E^x(z) + i\tilde Y(x)\tilde E^y(z) +i{\cal E}'(z)
\end{equation}
and its action on a node state is
\begin{equation}
\tilde K_+|m,n\rangle=({\cal
D}-m-n)\,|m,n\rangle+m\,|m+1,n\rangle+n\,|m,n+1\rangle.
\end{equation}
The equation for the coefficients is now
\begin{equation}\label{41}
({\cal D}-m-n)\,a_{m,n}+(m-1)\,a_{m-1,n}+(n-1)\,a_{m,n-1}=0.
\end{equation}
In the result for a given ${\cal D}$ we obtain the same type of
function as in the foregoing case with the same coefficients
$a_{-m,-n}=a_{m,n}$ as the corresponding coefficients for positive
$m$ and $n$, explicitly
\begin{equation}
a_{m,n}=(-1)^{m+n}\left(\begin{array}{c}-m_{\rm min}\\
-m\end{array}\right)\left(\begin{array}{c}-n_{\rm min}\\ -n
\end{array}\right).
\end{equation}

\noindent{\bf Case 3}. $m\geq0$, $n\leq0$: To obtain an action of
the Killing constraint ``towards the axes", $XE^x$ is promoted to an
operator according to (\ref{E1}) and (\ref{U1}) and $YE^y$ according
to (\ref{tilde}) and (\ref{U2}). In this way we obtain again
solutions with a finite number of nonzero coefficients. The equation
for the coefficients $a_{m,n}$ is the following
\begin{equation}
({\cal D}+m-n)\,a_{m,n}-(m+1)\,a_{m+1,n}+(n-1)\,a_{m,n-1}=0,
\end{equation}
their general form is
\begin{equation}
a_{m,n}=(-1)^{m-n}\left(\begin{array}{c}m_{\rm max}\\ m
\end{array}\right)\left(\begin{array}{c}-n_{\rm min}\\ -n
\end{array}\right).
\end{equation}
Now for each ${\cal D}<0$ the location of nonzero coefficients in
the fourth quadrant of the $(m,n)$ plane is bounded by the relation
$m_{\rm max}-n_{\rm min}=-{\cal D}$.\\

\noindent{\bf Case 4}. $m\leq 0$, $n\geq0$: This case is analogous
to the foregoing one with the roles of $m$ and $n$ exchanged and the
solution lying in the second quadrant. In all four cases the
coefficients can be normalized according to
\begin{equation}
\bar a_{m,n}=(-1)^{|m|+|n|}\left[\left(\begin{array}{c}2|m_{\rm m}|\\
|m_{\rm m}|\end{array}\right)\left(\begin{array}{c}2|n_{\rm m}|\\
|n_{\rm
m}|\end{array}\right)\right]^{-\frac{1}{2}}\left(\begin{array}{c}
|m_{\rm m}|\\ |m|\end{array}\right)\left(\begin{array}{c} |n_{\rm
m}|\\ |n|\end{array}\right),
\end{equation}
where $m_{\rm m}$ and $n_{\rm m}$ mean the $m$ or $n$ with the
maximal absolute value.

In dependence on the sign of point holonomy labels the $X$ operator,
as applied in the above four cases, can be written in the unified
form
\begin{equation}
\widehat{\!\widehat X}={\rm sign}(m)\left(\hat U_x^{\;{\rm
sign}(m)}-\bf1\right),
\end{equation}
acting as $\hat X$ or $\tilde X$, according to the sign of $m$. In
the next subsection we will introduce in the same way two versions
for link operators in dependence on the link label $k$, so we may
summarize the unified definitions. We write $A$ for $X$, $Y$, $\cal
A$ and $U_A$ for the corresponding holonomies, $\alpha$ for the
labels $m$, $n$, $k$ of a state function, $E^A$ for the conjugate
momenta. Then in general the following operators may be defined
\begin{equation}\label{26}
A\rightarrow{\rm sign}(\alpha)\left(U_A^{{\rm
sign}(\alpha)}-{\bf1}\right),\hspace{1cm} E^A\rightarrow
-iU_A^{-{\rm sign}(\alpha)}\,\frac{\delta}{\delta A}.
\end{equation}

So far we have presented four independent solutions to the
right-moving unidirectionality constraint, one in each quadrant. At
this stage, we can seemingly either restrict ourselves to solutions
in one quadrant with one version of the operator, or take together
two or all four kinds of solutions. Whether or not one of the latter
versions is necessary, depends in the end on the Hamiltonian
constraint operator, which determines the dynamics. However, at the
kinematical level we did not yet consider the case ${\cal D}>0$, and
a discussion of this also involves at least two solutions in two
different $(m,n)$ quadrants.

In all four cases considered above the eigenvalues of $\hat
K_1/\tilde K_1$ are positive. As already mentioned, in classical
terms $K_1$ represents the time derivative $\dot{\cal E}$, and for
right-moving waves, where $\dot{\cal E}=-{\cal E}'$, we have so far
obtained only solutions with $\dot{\cal E}\geq0$ and ${\cal
E}'\leq0$ at every node. When ${\cal E}'>0$ and the wave is going to
the right, $\dot{\cal E}$ must necessarily be negative.

Technically this can be achieved by choosing for the case ${\cal
E}'>0$ the quadrant $m\leq0$, $n\leq0$ and replace $\cal D$ by
$-\cal D$ in eq. (\ref{41}), whereas the first quadrant remains
reserved to ${\cal E}'<0$. In this way ${\cal E}'$ is changed to
$-{\cal E}'$, so that the right-moving constraint becomes
\begin{equation}\label{K2}
K_+=XE^x+YE^y-{\cal E}',
\end{equation}
which looks formally like $K_-$ in the original definition
(\ref{K}), but as now $\cal E$ goes to $-\cal E$, the meaning of
$K_1$ is now $-\dot{\cal E}$, and $\dot{\cal E}=-{\cal E}'$ again,
with $\dot{\cal E}<0$ and ${\cal E}'>0$. Rephrasing it in a
different way, for ${\cal E}'>0$ we have constructed a solution with
opposite orientation, moving to the left and backwards in time, and
reinterpret it as right-moving forward in time.

Effectively we redefine $K_+$ to be given by (\ref{K}) in the
quadrant $m>0$, $n>0$ for ${\cal E}'<0$, and by (\ref{K2}) in the
quadrant $m<0$, $n<0$ for ${\cal E}'>0$. This approach makes use of
two quadrants of the $(m,n)$ plane, but one could also find more
extended definitions involving all four quadrants. The occurrence of
different signs of triad components, leading to sectors of the
theory with different spatial triad orientations, is common in LQG
and loop quantum cosmology, see, for example \cite{LQC,kie}.

\subsection{Link operators and functions}
So far ${\cal D}=i{\cal E}'$ has been considered simply as an
integer number in order to match the integers $m$ and $n$. But, to
be consistent with the foregoing, we must also replace the classical
canonical pair $({\cal A},\cal E)$ by a pair of canonically
commuting operators, according to (\ref{calE}) and (\ref{25}), in
dependence of the sign of the link label $k$.

A single link holonomy $|k\rangle$ is not an eigenstate of
$\hat{\cal E}$ or $\tilde{\cal E}$. As $\hat{\cal E}$ is basically a
lowering operator, to be applied for $k>0$, and $\tilde{\cal E}$ is
a raising operator for $k<0$, eigenstates are in both cases of the
type of coherent states in the form of
\begin{equation}
|\kappa\rangle \sim \sum_{k=0}^\infty\frac{\kappa^k}{k!}\,|k\rangle
\hspace{1cm}\mbox{with}\hspace{1cm}\hat{\cal
E}\,|\kappa\rangle=-i\kappa\,|\kappa-1\rangle
\end{equation} for $\kappa>0$ and
\begin{equation}
|\kappa\rangle \sim
\sum_{k=0}^{-\infty}\frac{\kappa^{-k}}{(-k)!}\,|k\rangle
\hspace{1cm}\mbox{with}\hspace{1cm}\tilde{\cal
E}\,|\kappa\rangle=-i\kappa\,|\kappa+1\rangle
\end{equation}
for $\kappa<0$. Then the operator $\hat K_+$ in the version
(\ref{31}) ($\bar{\cal E}$ denotes the $\cal E$ operator for both
$\kappa>0$ and $\kappa<0$)
\begin{equation}
\hat K_+=\hat K_1 + i\bar{\cal E}_+-i\bar{\cal E}_-
\end{equation}
with $\bar{\cal E}_\pm$ acting on the links right and left from the
considered node. On a state $|\psi\rangle$ containing an eigenstate
of $\hat K_1$ with eigenvalue $\cal D$ (\ref{D}), as well as
eigenstates $|\kappa_\pm\rangle$ of $\bar{\cal E}_+$ and $\hat{\cal
E}_-$,
\begin{equation}
|\psi\rangle= \ldots|\kappa_-\rangle\otimes |{\cal
D}\rangle\otimes|\kappa_+\rangle\ldots,
\end{equation}
$\hat K_+$ acts in the way
\begin{equation}\label{ei}
\hat K_+|\psi\rangle=({\cal D}+\kappa_+-\kappa_-)\,|\psi\rangle
\end{equation}
and and for ${\cal D}=\kappa_--\kappa_+$ we have a solution of the constraint.\\

Eigenstates of $\hat{\cal E}$ have the following normalization
\begin{equation}
\langle\kappa|\kappa\rangle=\sum_{k=0}^\infty\frac{\kappa^{2k}}{(k!)^2}=I_0(2\kappa).
\end{equation}
$I_n(x)=(-i)^n\,J_n(ix)$ are modified Bessel functions. So the
normalized eigenfunctions are
\begin{equation}
|\kappa\rangle=\frac{1}{\sqrt{I_0(2\kappa)}}\,\sum_{k=0}^\infty\,\frac{\kappa^k}{k!}\,|k\rangle.
\end{equation}
With this normalization the expectation value of a positive $k$ in
an eigenstate becomes
\begin{equation}
\langle k\rangle=\kappa\,\frac{I_1(2\kappa)}{I_0(2\kappa)},
\end{equation}
with the fluctuation
\begin{equation}
\Delta
k=\kappa\,\sqrt{1-\left(\frac{I_1(2\kappa)}{I_0(2\kappa)}\right)^2}.
\end{equation}
For growing $\kappa$ the area expectation value $\langle k\rangle$
quickly approaches $\kappa$, whereas the area fluctuation $\Delta k$
grows only slowly (for example $\Delta k \approx 22$ for
$\kappa=1000$.) This is in accordance with the fact that for weak
gravitational waves transversal area variations are small.

\section{Conclusion}

In the present approach we have found well-behaved kinematical
solutions to the unidirectionality constraint for plane
gravitational waves. For complete solutions we need two sectors of
the theory with different signs of triad variables. These different
signs distinguish different orientations of space. For physical
reasons one may expect the action of quantum operators on states
with only sign differences to act in a very closely related way, as
operator pairs like $(\hat X,\tilde X)$, $(\hat E^x,\tilde E^x)$ or
$(\hat K_1,\tilde K_1)$ (defined in (\ref{hatk1}) and (\ref{K1t}))
do in their respective domains.

Even if from the mathematical point of view the approach with
canonically commuting operators may appear less natural, physically
it yields substantially better results than the previous one in
\cite{adel}, where more common methods were used. After all, with
the aid of step and sign functions and $\hat K_1$ and $\tilde K_1$
it is possible to formulate the right-moving constraint operator
constructed in this paper in a closed form, when the link functions
are coherent states:
\begin{eqnarray}
&&\widehat{\widehat K}_+:=\Theta(m)\Theta(n)\hat
K_1+\Theta(-m)\Theta(-n)\tilde
K_1+\\[2mm]
&&{\rm sign}(\kappa_+-\kappa_-)[\Theta(k_+)\hat{\cal
E}_++\Theta(-k_+)\tilde{\cal E}_+-\Theta(k_-)\hat{\cal
E}_--\Theta(-k_-)\tilde{\cal E}_-]\nonumber.
\end{eqnarray}
Whether or not dynamical solutions can be of the considered type, or
whether all four quadrants of $(m,n)$ are needed for a consistent
dynamics of the model, is expected to be determined by the action of
the Hamiltonian constraint on the states found in this paper. This
problem will be the subject of future work.

It also turned out that at the kinematical level the Minkowski
vacuum cannot be modeled by solutions of both the quantum
constraints corresponding to the classical constraints $K_1=0$ and
${\cal E}'=0$. Assuming globally ${\cal E}={\rm const}.$, we are
left with the zero volume and zero length state $|0,0\rangle$ at
every node, effectively the same as a state without nodes at all.

Now, as strictly constant cross section area along the $z$ axis is
impossible, there must be small, but nonvanishing spatial
fluctuations in $\cal E$. In the sequel, area fluctuations lead to
volume and length fluctuations, as it follows from the calculations
in this paper. This indicates that fluctuations have the form of
small left- or right-moving waves, there are no static fluctuations
at nodes, while ${\cal E}'$ would be zero. Again, what these
fluctuations are like in a realistic dynamical model is a matter of
the Hamiltonian constraint. In a dynamical Minkowski space solution
we can expect a balanced mixture of right- and left-moving
fluctuations everywhere along the $z$ axis.

\end{document}